\def\n{\noindent}
\def\SCC{{\it Solid State Commun.}\ }

\def\P{$P$}

  \def\ket{\vert \vert  \{ \emptyset \} \rangle}
  \def\ket2{\vert \vert \otimes \{ R \} \rangle}

\def\.#1{\mathaccent 95#1}
\def\^#1{\mathaccent 94 #1}
\def\~#1{\mathaccent "7E #1}

\def\unit{{\cal I}}

  \def\ket{\vert \vert  \{ \emptyset \} \rangle}
  \def\ket2{\vert \vert \otimes \{ R \} \rangle}

\def\be{\begin{equation}}
\def\ee{\end{equation}}

\documentclass[12pt,epsfig]{iopart}
\usepackage{epsfig}
\begin{document}
\setcounter{page}{1}
\title{Magnetic properties of X-Pt (X=Fe,Co,Ni) alloy systems}
\author{\bf Durga Paudyal, Tanusri Saha-Dasgupta and Abhijit Mookerjee}
\address { S.N. Bose National Centre for Basic Sciences,
JD Block, Sector 3, Salt Lake City, Kolkata 700098, India \\ 
Email: dpaudyal@bose.res.in, tanusri@bose.res.in, abhijit@bose.res.in}
\date{\today}
\begin{abstract} 
We have studied the electronic and magnetic properties of Fe-Pt, Co-Pt 
and Ni-Pt alloy systems in ordered and disordered phases. The influence 
of various exchange-correlation functionals on values of equilibrium 
lattice parameters and magnetic moments in ordered Fe-Pt, Co-Pt 
and Ni-Pt alloys have been studied using linearized muffin-tin orbital 
method. The electronic structure calculations for the disordered alloys
have been carried out using augmented space 
recursion technique in the framework of tight binding linearized muffin-tin 
orbital method. The effect of short range order has also been studied in the 
disordered phase of these systems. The results show good agreements with 
available experimental values. 
\end{abstract}

\pacs{71.20,71.20c }

\section{Introduction}
The magnetic and chemical interactions in solid solutions, their
interdependence and the role they play in determining the electronic
and magnetic properties of  transition metal alloys have been the subject 
of extensive research since many years. The interplay between magnetism and 
spatial order in transition metal alloy systems has been  extensively studied  
both experimentally  \cite{kn:cad},\cite{kn:mir}-\cite{kn:cad3}
and using  phenomenological models based on statistical thermodynamics
\cite{kn:sato}-\cite{kn:hen},\cite{kn:jac}-\cite{kn:kak}.

In this communication, we studied the electronic and magnetic properties
of ordered as well as disordered phase of the Fe-Pt, Co-Pt and Ni-Pt.
Many studies on optical and magneto-optical characterization of these systems
are available in recent literatures \cite{kn:uba}. 
Nevertheless, a systematic first-principles
study bringing out the interdependence of the magnetic and chemical ordering and the
trend in this alloy series is lacking. The present communication aims at a 
systematic and comparative 
first principles study of the electronic structure and magnetism in these systems, 
using techniques based on the local spin density approximation (LSDA) of the density 
functional theory. 

Considering the case of ordered alloys, we have carried out a thorough study including 
careful investigation of  the influence
of various local as well as non-local exchange correlation functionals on the 
value of the equilibrium  lattice parameters and magnetic moments of ordered Fe-Pt, 
Co-Pt and Ni-Pt alloy systems. 

The calculational scheme used for our calculations of disordered alloys is based on
augmented space recursion (ASR) technique. While the majority of the existing electronic 
structure calculations on {\sl disordered} alloys have been  based on the coherent potential 
approximation (CPA), the CPA, being a single-site mean-field approximation, cannot take into 
account the effect, at a site, of its immediate environment. As an alternative approach, Saha 
\etal \cite{kn:asr} have introduced the augmented space recursion (ASR) based on the combination 
of the augmented space formalism (ASF) first suggested by Mookerjee \cite{kn:mook} and the 
recursion method of Haydock \etal \cite{kn:hhk}. In this formalism, the configuration averaging 
is carried out without having to resort to single-site approximations. The recursion method allows 
one to take into account the effect of the local environment on electronic properties. Moreover, 
the convergence of various physical quantities calculated through recursion with the number
of recursion steps and subsequent termination has been studied in great detail 
\cite{kn:vol35,kn:sm}. Among the various advantages of the ASR in going beyond the single-site 
approximation is the possibility of inclusion of local lattice distortions \cite{kn:latdis} 
which is important in the case of alloys with size mismatch between components as in 
the case of Fe-Pt, Co-Pt and Ni-Pt. 

An important aspect in understanding the interplay between magnetism and ordering in disordered 
transition metal alloys involves investigation of the influence of local environment, namely the 
short-range ordering
(SRO) effect, on electronic and magnetic properties of these alloys.
There have been determination of  SRO parameters for different degrees of disorder 
using first principles techniques \cite{kn:zun,kn:staun,kn:john} or extraction of these  
parameters from experiments and analysis of their effect on electronic
structure and properties \cite{kn:borici,kn:zun2,kn:abri}.
SRO for a disordered binary alloy A$_x$B$_{1-x}$ is described, for example, 
by the Warren-Cowley parameter \cite{kn:cow} which is defined as :

\begin{equation}
\alpha_{r}^{AB} = 1-\frac{P_{r}^{AB}}{y} \nonumber
\end{equation}

\n for the B atom occupying the r-th nearest neighbour site of the central A atom.
$y=1-x$ denotes the macroscopic
concentration of species $B$ and $P_{r}^{AB}$ is the joint probability of
finding a $B$ atom anywhere in the $r^{th}$ shell.  

Mookerjee and Prasad \cite{kn:as} introduced a method for calculating the electronic structure of 
disordered alloys with short range order (SRO) which is based on a generalization of the  augmented 
space theorem \cite{kn:mook}. Saha \etal \cite{kn:saha} implemented this within the framework of 
recursion method.  
Later  Ghosh \etal \cite{kn:gcsm} extended the technique to  magnetic 
Co-Pt and Co-Pd systems. In present paper we have carried out charge-self-consistent 
calculations based on this generalized ASR technique to examine the short range ordering 
effect in Fe-Pt, Co-Pt and Ni-Pt systems.

The paper has been organized in the following manner. Section 2 is devoted to theoretical
and computational details. The results of our study
along-with comparison with existing experimental and theoretical studies have been discussed
in section 3. We end the paper with the summary and conclusion in section 4. Some of the relevent 
equations of generalized augmented space recursion method have been put in the appendix.    

\section{ Theoretical and Computational details}
For ordered structures we have performed the total energy density functional calculations. 
The Kohn-Sham equations were solved in the local spin density approximation (LSDA) with von 
Barth-Hedin (vBH) \cite{kn:vbh} and Vosko-Wilk-Nusair (VWN) \cite{kn:vwn} exchange correlations as well as in the 
non-local (generalized 
gradient approximation (GGA)) Langreth-Mehl-Hu (LMH) \cite{kn:lmh} and Perdew-Wang exchange (PW) \cite{kn:pw} 
correlations. The calculations have been performed in the basis of tight binding linear muffin-tin orbitals in the 
atomic sphere approximation (TB-LMTO-ASA) \cite{kn:ajs}-\cite{kn:ddsam} including 
combined corrections. The calculations are semi-relativistic through inclusion of mass-velocity and 
Darwin correction terms. The k-space integration was carried out with 32$\times$32$\times$32 mesh 
resulting 969 k points for cubic 
primitive structures and 2601 k points for tetragonal primitive structures in the 
irreducible part of the corresponding Brillouin zone. The convergence of the magnetic moments with respect 
to k-points have been checked. To have theoretical estimates of the equilibrium lattice parameters, 
we have carried out the minimization of the self-consistent TB-LMTO-ASA total energies varying lattice 
parameters for Fe-Pt, Co-Pt and Ni-Pt alloys at different concentrations. 

Our disordered calculations are based on the generalized ASR technique \cite{kn:as}-\cite{kn:gcsm},\cite{kn:bs}-\cite{kn:ppb}.
The Hamiltonian in the TB-LMTO minimal basis is sparse and therefore suitable for the
application of the recursion method introduced by Haydock \etal \cite{kn:hhk}.
The ASR allows us to calculate the configuration averaged Green functions.
It does so by augmenting the Hilbert space spanned by the TB-LMTO basis
by the configuration space of the random Hamiltonian parameters. The configuration average is expressed {\sl exactly}
as a matrix element in the augmented space.  A generalized form of this methodology is capable of
taking into account the effect of short range order. Please see appendix for relevant equations. 
The initial guess TB-LMTO potential parameters for the self-consistency iterations
for disordered alloy calculations are taken to be the potential parameters of pure constituents.
In subsequent iterations
the potential parameters are obtained from the solution of the Kohn-Sham equation

\begin{equation}
\left\{ -\frac{\hbar^{2}}{2m} \nabla^{2} + V^{\nu\sigma} - E\right\} \phi^{\nu}_{\sigma}(r_{R}, E)
\; =\; 0 \end{equation}

where,

\begin{equation}
V^{\lambda\sigma}(r_{R})\;  = \; V_{core}^{\lambda\sigma}(r_{R}) + V_{har}^{\lambda\sigma}(r_{R})
                           + V_{xc}^{\lambda\sigma}(r_{R}) + V_{mad}
\end{equation}

\noindent The electronic position within the atomic sphere centered at $R$ is given by $r_{R}$ =$r-R$.
$\sigma$ is the spin component. The core potentials are obtained from atomic calculations and are available for
most atoms.
For the treatment of the Madelung potential, we follow the procedure suggested by Kudrnovsk\'y \etal \cite{kn:kd}
and use an extension of the procedure proposed by Andersen \etal
\cite{kn:ajs}.
 We choose the atomic sphere
radii of the components in such a way that they preserve the total volume on the
average and the individual atomic spheres are almost charge neutral. This ensures
that total charge is conserved,  but each atomic sphere carries no excess
charge. In doing so, one needs to be careful about the sphere overlap which should be under
certain limit so as to not violate the
atomic sphere approximation.

In these calculations one also needs to be very careful about the convergence of Fermi energy  as well 
as that of magnetic moments. In fact, errors can arise in the augmented space recursion because one can 
carry out only finite number of recursion steps and then terminate the continued fraction using available 
terminators. Also one chooses a large but finite part of the augmented space nearest neighbour map and
ignores the part of the augmented space very far from the starting state. This is also a source of error.

The formulation of the augmented space recursion as described in appendix and used for the calculation in the 
present paper is the energy dependent augmented space recursion in which the disordered Hamiltonian 
with diagonal as well as off-diagonal disorder is recast into an energy dependent Hamiltonian having only diagonal disorder.
We have chosen a few seed points across the energy spectrum uniformly, carried out recursion on those points
and spline fit the coefficients of recursion through out the whole spectrum. This enabled us to carry out
large number of recursion steps since the configuration space grows significantly less faster for diagonal
as compared with off diagonal disorder. Convergence of physical quantities with recursion steps have been 
discussed in detail earlier by Ghosh \etal \cite{kn:gdm,kn:sdgthesis}.

We have checked  the convergence of Fermi energy and magnetic moments with respect to
recursion steps and the number of seed energy points for the case of NiPt$_3$ system.
We have found that the Fermi energy and magnetic moments converge beyond seven recursion 
steps and thirty five seed energy points. In our all calculations
reported in the following have been carried out with eight recursion steps
and thirty five seed energy points.

\begin{table}
\caption{The equilibrium lattice parameters in a.u. of FePt, CoPt and NiPt systems in ordered structures
with various choices of exchange correlation functionals. See text for various abbreviations.}
\begin{center}
\begin{tabular}{|c|c|c|c|c|c|}
\br
x     & vBH & VWN & LMH & PW & Expt.          \\ \mr
\multicolumn{6}{|c|}{\bf Fe$_{1-x}$Pt$_{x}$}\\
\mr
0.00~(BCC)  & 5.28     & 5.30     & 5.36     & 5.54     & 5.406 \cite{kn:pearson}     \\
~~~~~(FCC)  & 6.47     & 6.47     & 6.53     & 6.63     & 6.877 \cite{kn:pearson}     \\
0.25(L1$_2$)& 6.71     & 6.91     & 6.99     & 7.21     & 7.049 \cite{kn:kashyap}     \\
0.50(L1$_0$)& a = 7.16 & a = 7.18 & a = 7.22 & a = 7.46 & a = 7.253 \cite{kn:pearson}\\
            & c = 6.94 & c = 6.94 & c = 7.02 & c = 7.26 & c = 7.020 \cite{kn:pearson}\\
0.75(L1$_2$)& 7.25     & 7.27     & 7.30     & 7.54     & 7.313 \cite{kn:pod} \\ \mr
\multicolumn{6}{|c|}{\bf Co$_{1-x}$Pt$_{x}$}\\
\mr
0.00~(hex)  & a = 4.65 & a = 4.66 & a = 4.70 & 4.83     & 4.728 \cite{kn:pearson}\\
            & c = 7.48 & c = 7.49 & c = 7.59 & 7.78     & 7.675 \cite{kn:pearson}\\
~~~~~(FCC)  & 6.55     & 6.56     & 6.63     & 6.81     & 6.684 \cite{kn:pearson}\\
0.25(L1$_2$)& 6.78     & 6.80     & 6.86     & 7.06     & 6.923 \cite{kn:kashyap}\\
0.50(L1$_0$)& a = 7.14 & a = 7.14 & a = 7.18 & a = 7.40 & a = 7.204 \cite{kn:pearson}\\
            & c = 6.78 & c = 6.78 & c = 6.86 & c = 7.08 & c = 7.007 \cite{kn:pearson}\\
0.75(L1$_2$)& 7.21     & 7.22     & 7.25     & 7.50     & 7.240 \cite{kn:pearson}\\ \mr
\multicolumn{6}{|c|}{\bf Ni$_{1-x}$Pt$_{x}$}\\
\mr
0.00~(FCC)  & 6.54     & 6.55     & 6.61     & 6.80     & 6.646 \cite{kn:pearson} \\
0.25(L1$_2$)& 6.77     & 6.78     & 6.84     & 7.05     & 6.890 \cite{kn:pisanty} \\
0.50(L1$_0$)& a = 7.16 & a = 7.16 & a = 7.18 & a = 7.42 & a = 7.209 \cite{kn:pearson}  \\
            & c = 6.63 & c = 6.64 & c = 6.74 & c = 6.96 & c = 6.769 \cite{kn:pearson}  \\
0.75(L1$_2$)& 7.20     & 7.21     & 7.24     & 7.49     & 7.251 \cite{kn:pisanty} \\
1.00~(FCC)  & 7.37     & 7.38     & 7.40     & 7.66     & 7.400 \cite{kn:pearson} \\ \br
\end{tabular}
\end{center}
\label{tab1}
\end{table}

\section{Results and discussions}
\subsection{Lattice Parameters}

In Table 1, we quote the values of equilibrium lattice parameters, obtained 
by minimizing the total energy with respect to the lattice parameters for L1$_2$ superstructures at 25 and 75$\%$ 
and L1$_0$ superstructure at 50$\%$ concentration of Pt in 
Fe-Pt, Co-Pt and Ni-Pt alloy systems with different 
choice of local as well as non-local exchange 
correlation potentials. The first comment is that non-local exchange correlation potentials seem to decrease
overbinding and predict larger equilibrium lattice parameters than the local ones. The PW
seems to go overboard and give estimates of the equilibrium lattice parameters which are {\sl larger}
than the experimental values. The best agreement with experiment is found to be LMH. 

\begin{table}
\caption{The local and average magnetic moments of Fe-Pt system in ordered structures with various 
choices of exchange correlation functionals.}
\begin{center}
\begin{tabular}{|c|l|c|c|c|c|c|c|}
\br
concentration& XC used/Expt/ & \multicolumn{6}{|c|}{magnetic moment ($\mu_B$/atom) of}\\ \cline{3-8}
of Pt        & Ref.          &\multicolumn{3}{c|}{with eq. lat. par.} & \multicolumn{3}{c|}{with expt. lat. par.}      \\ \cline{3-8}
             &               & Fe     & Pt   &average & Fe     & Pt   &average\\ \cline{1-8}
\mr
0.00(BCC)&vBH(this work)& 2.15          &               & & 2.25 & &     \\
         &VWN(this work)& 2.21          &               & & 2.30 & &     \\
         &LMH(this work)& 2.29          &               & & 2.33 & &     \\
         &PW(this work) & 2.55          &               & & 2.35 & &     \\
\cline{2-8}
         & Expt. \cite{kn:crc}        &                &               & & 2.22  & &     \\
\mr
0.25(L1$_2$)&vBH(this work)& 0.00          & 0.00         & 0.00   & 2.57           & 0.32         & 2.01 \\
            &VWN(this work)& 2.46          & 0.29         & 1.92   & 2.64           & 0.34         & 2.06 \\
            &LMH(this work)& 2.63          & 0.33         & 2.06   & 2.70           & 0.35         & 2.11 \\
            &PW(this work) & 2.78          & 0.35         & 2.17   & 2.68           & 0.37         & 2.10\\
            & Auluck \etal (vBH) \cite{kn:kashyap}&&&     & 2.56   & 0.26           & 1.99 \\
            & Podgorny (VWN) \cite{kn:pod} & 2.51         & 0.26   & 1.95        & & & \\
            & Hasegawa \cite{kn:has}  &&&  & 2.50         & 0.50   & 2.0  \\
\cline{2-8}
            & Expt. \cite{kn:kashyap}   &&&& 2.70         & 0.50   & 2.15\\
\mr
0.50(L1$_0$)&vBH(this work)& 2.73          & 0.35         & 1.54 & 2.81          & 0.35         & 1.58 \\
            &VWN(this work)& 2.79          & 0.35         & 1.57 & 2.85          & 0.35         & 1.60 \\
            &LMH(this work)& 2.88          & 0.35         & 1.61 & 2.90          & 0.35         & 1.63 \\
            &PW(this work) & 3.01          & 0.36         & 1.69 & 2.86          & 0.36         & 1.61 \\
            & Osterloh \etal \cite{kn:ost}&&&             & 2.92 & 0.38         &                  \\
            & Podgorny (VWN) \cite{kn:pod} & 2.85         & 0.30 & 1.57 & & & \\
\cline{2-8}
             & [Expt.] \cite{kn:ost}& &               & & 2.80 & & 0.77 \\
\mr
& \multicolumn{7}{|c|}{Ferromagnetic calculation}\\ \cline{2-8}
            &vBH(this work)& 2.99          & 0.31           & 0.98          & 3.10  & 0.32         & 1.02 \\
            &VWN(this work)& 3.12          & 0.32           & 1.02          & 3.15  & 0.33         & 1.03 \\
            &LMH(this work)& 3.19          & 0.34           & 1.05          & 3.20  & 0.34         & 1.06 \\
            &PW(this work) & 3.24          & 0.39           & 1.11          & 3.12  & 0.37         & 1.06 \\
             & Podgorny \cite{kn:pod}      & 3.22           & 0.34          & 1.06  & & & \\
0.75(L1$_2$) & Tohyama \etal \cite{kn:toh}&&&                                     & 4.21 & 0.33           &\\
\cline{2-8}
& \multicolumn{7}{|c|}{Anti-ferromagnetic calculation}\\ \cline{2-8}
            &vBH(this work)& 3.11         & 0.15           &               & 3.16                 & 0.15 & \\
            &VWN(this work)& 3.17         & 0.15           &               & 3.20                 & 0.15 & \\
            &LMH(this work)& 3.24         & 0.15           &               & 3.25                 & 0.16 & \\
            &PW(this work) & 3.31         & 0.17           &               & 3.18                 & 0.16 & \\
            & Podgorny \cite{kn:pod}& 3.46 & 0.16 & & & &\\
            & Tohyama \etal \cite{kn:toh} & & &  & 4.13  & 0.00 &\\
\cline{2-8}
             & [Expt.]\cite{kn:kky}  & &                &               &3.3&& \\
\br
\end{tabular}
\end{center}
\label{tab2}
\end{table}

\subsection{Magnetism of Fe-Pt Alloys}

\subsubsection{Ordered Alloys :}

In Table 2, we show two sets of calculations for magnetic moments in ordered Fe-Pt alloys. 
In first set of calculations, we have calculated local as well as average magnetic 
moments corresponding to the theoretically estimated lattice parameters obtained via energy minimization 
procedure. In second set, calculations were done using 
experimental lattice parameters. 

For Fe$_3$Pt alloy in L1$_2$ super-structure the use of non-local exchange correlation
functionals LMH appear to give  better agreement with experimental values \cite{kn:kashyap} for local and
average magnetic moments as compared to local exchange correlation functionals.
This holds good for both the choices of lattice parameters. 
The results for average and local magnetic moments from previous works by Auluck \etal \cite{kn:kashyap} 
and Podgorny \cite{kn:pod}, both using TB-LMTO,  are in agreement with our corresponding 
results as can be seen from Table 2. The differences seen with these results are primarily due to 
different computational details.
Auluck \etal \cite{kn:kashyap} and Podgorny \cite{kn:pod} have used frozen core approximation in their calculations 
without taking into account of $f$ states for Pt.
Podgorny and Auluck \etal in their calculations used 286 and 84 k points 
in the irreducible part of the Brillouin Zone (BZ) respectively. On the other hand, our 
calculations are all electron 
calculations taking a $spdf$ minimal basis for Pt and using 969 k points in irreducible part of 
BZ. The local magnetic moment on Pt sites obtained by Hasegawa \etal \cite{kn:has} using augmented plane 
wave (APW) method is in exact agreement to the corresponding experimental value though their 
average magnetic moment and local magnetic moment on Fe sites are lower (by 0.20 $\mu_B$ for average 
and 0.15 $\mu_B$ for Fe sites) than the corresponding 
experimental estimates \cite{kn:kashyap}. Our calculation 
using the  vBH functional for the exchange correlation
potential and theoretically estimated lattice parameter leads to the conclusion of a  non-magnetic ground 
state which is in agreement with that found in a previous study by Kubler \etal \cite{kn:kub}. This once
again emphasizes that magnetic moments are very sensitively dependent on the particular exchange-correlation
functional used and the detailed accuracy of the numerical calculations.

For FePt alloys the local magnetic moment of Fe site in L1$_0$ superstructure
calculated using vBH exchange correlation potential and experimental 
lattice parameter shows closest agreement with experimental value
\cite{kn:ost}. The LMH based estimates of the local magnetic moment on Fe sites
 are rather large as compared with the one experimental datum available \cite{kn:ost}.
The experimental value for local magnetic moment of Pt in this concentration is not available. The 
experimentally estimated average magnetic moment is significantly lower than that of the calculated values 
using both local as well as non local exchange correlations. However all the available theoretical estimates 
by different groups \cite{kn:pod,kn:ost} are significantly high, just like ours as compared to 
the experimental estimate quoted by Osterloh \cite{kn:ost}. The experimental result may be interpreted assuming 
the magnetic moment at the Fe and Pt sites to be arranged antiparallely giving rise to ferri-magnetic 
ground state. 
However, we were unable to show any  theoretical evidence for this and our  calculations 
do predict a stable ferromagnetic alignment as pointed out by Osterloh \etal \cite{kn:ost}. 
As in the case of Fe$_3$Pt, the slight difference between the values 
obtained by Podgorny \cite{kn:pod} and by us is again due to the difference in the calculational 
details. 
In addition to using frozen core approximation and neglect 
of f states in Pt site, Podgorny has assumed the cubic crystal structure for FePt in L1$_0$ 
structure while in reality it is tetragonal. In our calculations, we have assumed the experimentally observed 
tetragonal structure. The local magnetic moments obtained by Osterloh \etal \cite{kn:ost} 
using augmented spherical wave method are higher 
than ours as well as calculations by Podgorny \cite{kn:pod}.

The experimental ground state ordered 
magnetic phase FePt$_3$ is antiferromagnetic. We have carried out calculations on this alloy  both in
the ferromagnetic as well as the antiferromagnetic structures.
We have found the total energy in the case of antiferromagnetic structure is indeed lower than that of ferromagnetic structure.
In the ferromagnetic calculation, the local as well as average 
magnetic moment obtained by Podgorny \cite{kn:pod} using VWN exchange correlation potential with
theoretical estimates of lattice parameter is in close agreement with our corresponding value. 
The calculated local magnetic moment on Fe sites by Tohyama \etal \cite{kn:toh} 
using an empirical tight binding model is significantly higher 
than both ours and that  of Podgorny \cite{kn:pod}. 
Our calculated magnetic moment on Fe site for the antiferromagnetic structure 
using PW non local exchange correlation with theoretically estimated lattice 
parameter is in closest agreement with the experimental value \cite{kn:kky}. This is the one
case where LMH underestimates the staggered magnetization. 

\begin{figure}
\centering
\epsfxsize=4.0in\epsfysize=3.5in
\rotatebox{270}{\epsfbox{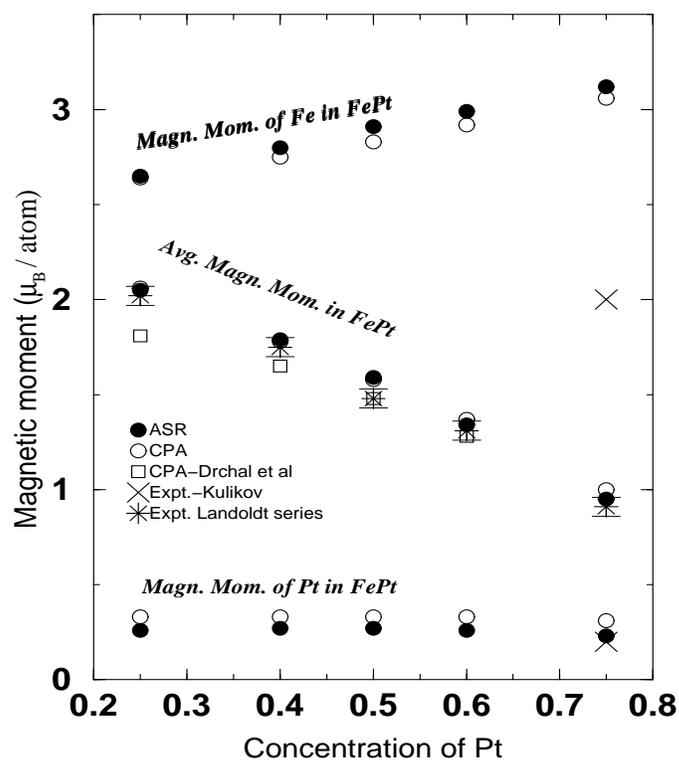}}
\caption{Magnetic moments in disordered Fe-Pt alloy systems using two different configuration averaging
methods namely augmented space
recursion (ASR) and coherent potential approximation (CPA) as compared to available experimental values given in
Landoldt series \cite{kn:expt}.}
\label{fig1}
\end{figure}

\begin{figure}
\centering
\epsfxsize=3.5in\epsfysize=3.5in
\rotatebox{270}{\epsfbox{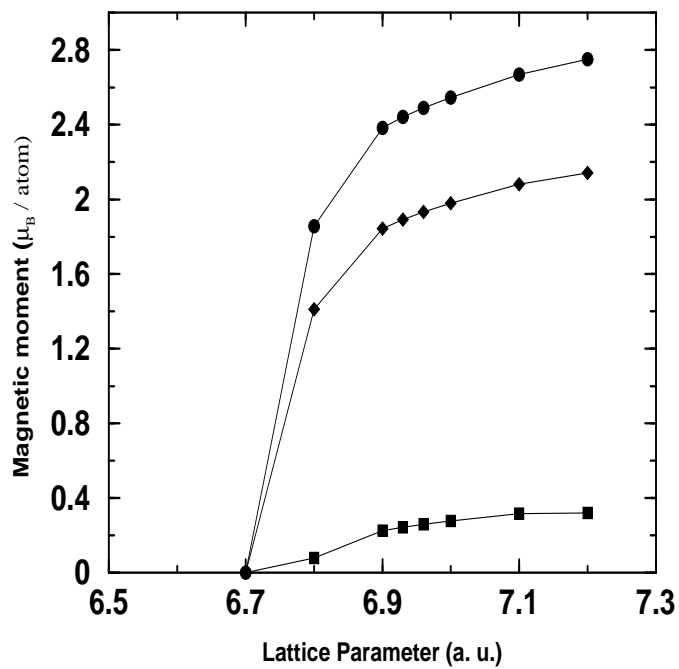}}
\caption{Magnetic moments as a function of lattice parameters for 25$\%$ concentration of Pt
in disordered Fe-Pt alloy. Circles, squares and diamonds denote local magnetic moment Fe site, local magnetic
moment on Pt site and average magnetic moment respectively.}
\label{fig2}
\end{figure}

\subsubsection{Disordered Alloys :}
 In Figure 1, we compare our calculated disordered magnetic moments using augmented space recursion
with the available experimental values taken from Landoldt series \cite{kn:expt} as well as with CPA
calculations. 
The average magnetic moments agree quite well with the corresponding experimental values in all concentrations. 
The numerical values of local as well as average magnetic moments calculated 
using LMTO-CPA 
are in agreement with those obtained using the ASR. This  shows that the single site approximation  
like CPA works well for the Fe-Pt disordered alloys.
The average magnetic moments obtained by Drchal \etal \cite{kn:hd} using CPA matches well for most  
concentrations though they deviate a  bit at low concentrations of Pt. 
Our calculations uses charge neutral spheres 
to reduce the effect of Madelung whereas Drchal \etal \cite{kn:hd} have used 
equal Weigner Seitz radii of both constituents and the effect of Madelung due to charge transfer was 
taken into account using screened impurity model \cite{kn:hd}. The local moment on the Fe sites increases
towards the isolated Fe moment as the concentration of Pt increases. This is an
indication of the fact that local environmental effects are unimportant and consequently the CPA and ASR results
agree closely.

In 25$\%$ concentration of Pt there is invar effect which shows anomalies in the thermal expansion. 
We have observed two minima of total energy one with a high moment and a large lattice constant 
6.93 au and  the other with a zero moment and small lattice constant 6.71 au. The total energy difference 
between the magnetic and non magnetic states is 2.4 mRyd/atom which is higher than the calculations by 
Drchal \etal (0.7 mRyd/atom) and lower than that of Staunton \etal (15.7mRyd/atom). In Figure 2, we 
show the behaviour of magnetic moment as a function of lattice parameter which shows non magnetic to 
ferromagnetic transition at 6.71 a.u. Our calculated average 
as well as local magnetic moment on Fe and Pt sites corresponding to theoretically estimated 
lattice parameter 6.93 a.u. via total energy minimization on magnetic state
are respectively 1.89, 2.44 and 0.24 $\mu_B$. 

Table 3 summarizes the known experimental and earlier theoretical results on disordered FePt with
25$\%$ Pt. 
The reported experimental results in this case differ to each other. The localized 
components of the magnetic moments for Fe (2.03 $\pm$ 0.02 $\mu_B$) and Pt (0.34$\pm$ 0.08 $\mu_B$) 
were estimated from spin polarized neutron diffraction measurements by Ito \etal \cite{kn:ito} 
 , while the magnetization 
measurements of Caporaletti and Graham \cite{kn:capo} indicated moments of 2.75 and 0.45 $\mu_B$ for Fe and Pt 
respectively. The values of average magnetic moments quoted in Landoldt series for different
experiments are 2.02 and 2.27 $\mu_B$. The theoretical estimates based on different methods also differ from
one another.
These differences are mainly due to 
the differences in the computational details chosen in each framework and also the 
approximations being used in each method.    

\begin{table}
\caption{Various estimates of the local and averaged magnetic moments in Bohr-magnetons for disordered Fe$_{75}$Pt$_{25}$ alloy.}
\centering
\begin{tabular}{|l|c|c|c|}\br
Author & Fe & Pt & Average \\ \mr
Expt. \cite{kn:ito} & 2.03$\pm$0.02 & 0.34$\pm$ 0.08 & 1.61$\pm$0.03\\ \mr
Expt. \cite{kn:capo} & 2.75 & 0.45 & 2.20 \\ \mr
Expt. (a) \cite{kn:expt} &      &      & 2.02 \\ \mr
Expt. (b) \cite{kn:expt} &      &      & 2.27 \\ \mr
LMTO-CPA \cite{kn:hd}          &    &      & 1.81 \\ \mr
KKR-CPA \cite{kn:stau}         &2.80     &  0.23    & 2.16 \\ \mr
LCAO-CPA \cite{kn:kvhe}        &     &      & 2.17 \\ \mr
ASR (this work)                & 2.44 & 0.24 & 1.89 \\ \br
\end{tabular}
\label{tab3}
\end{table}

For the 75$\%$ concentration of Pt our estimate of the magnetic moment on Fe sites is higher than that 
measured by  Kulikov \etal \cite{kn:kky} (which is about 2$\mu_B$). 
In order to check the possible short range order effect, we have checked the variation of total energy 
as a function of short range order and found that 
the total energy decreases as short range order goes from positive (segregation side) to 
negative (ordering side) confirming this system as an ordering system. 
We have also checked the variation of the magnetic moments as functions of the SRO parameter. We find
that both the local and average magnetic moments increase as the SRO parameter goes from the segregating
to the ordering side.  This is justified by the fact that the  
magnetic moment of Fe is enhanced when it is surrounded by Pt  as we have seen in the ordered alloys.
We therefore conclude that the discrepancy with the experimental data of Kulikov \etal \cite{kn:kky} 
can not be due to the short range ordering effect, probably the other possible factors influencing 
the experimental results need to be considered.

\begin{table}
\caption{The local and average magnetic moments of Co-Pt system in ordered structures with various 
choices of exchange correlation functionals.}
\begin{center}
\begin{tabular}{|c|l|c|c|c|c|c|c|}
\br
concentration& XC used/Expt/ & \multicolumn{6}{|c|}{magnetic moment ($\mu_B$/atom) of}\\ \cline{3-8}
of Pt        & Ref.          &\multicolumn{3}{c|}{with eq. lat. par.} & \multicolumn{3}{c|}{with expt. lat. par.}      \\ \cline{3-8}
             &               & Co     & Pt   &average & Co     & Pt   &average\\ \cline{1-8}
\mr
0.00(hex)&vBH(this work)& 1.55          &                &  & 1.60 &    &      \\
         &VWN(this work)& 1.58          &                &  & 1.62 &    &      \\
         &LMH(this work)& 1.62          &                &  & 1.64 &    &      \\
         &PW (this work)& 1.67          &                &  & 1.63 &    &      \\
\cline{2-8}
          & Expt. Kootte \cite{kn:kootte}&                &                &  & 1.58 &    &      \\
\mr
~~~~(FCC)&vBH(this work)& 1.57        &                &  & 1.62 &    &      \\
         &VWN(this work)& 1.60        &                &  & 1.64 &    &      \\
         &LMH(this work)& 1.65        &                &  & 1.67 &    &      \\
         &PW (this work)& 1.70        &                &  & 1.66 &    &      \\
\cline{2-8}
          & Expt. Kootte \cite{kn:kootte}     &              &                &  & 1.61  &    &      \\
\mr

0.25(L1$_2$)&vBH(this work)& 1.56          & 0.35         & 1.26 & 1.69          & 0.39         & 1.37 \\
            &VWN(this work)& 1.63          & 0.37         & 1.32 & 1.73          & 0.40         & 1.40 \\
            &LMH(this work)& 1.73          & 0.40         & 1.40 & 1.76          & 0.39         & 1.42 \\
            &PW(this work) & 1.80          & 0.39         & 1.45 & 1.74          & 0.41         & 1.41 \\
            & Auluck \etal\cite{kn:kashyap}&        &               &       & 1.39           & 0.38          & 1.14  \\
            & Kootte \cite{kn:kootte}&         &               &       & 1.64           & 0.36          & 1.32  \\
\mr
0.50(L1$_0$)&vBH(this work)& 1.69          & 0.38         & 1.03 & 1.79         & 0.38         & 1.09 \\
            &VWN(this work)& 1.74          & 0.39         & 1.07 & 1.83         & 0.39         & 1.11 \\
            &LMH(this work)& 1.82          & 0.40         & 1.11 & 1.87         & 0.39         & 1.13 \\
            &PW(this work) & 1.91          & 0.42         & 1.16 & 1.83         & 0.40         & 1.12 \\
            & Auluck \etal\cite{kn:kashyap}&        &               &       & 1.85           & 0.38          & 1.12 \\
            & Kootte \cite{kn:kootte}&        &               &        & 1.69           & 0.37          & 1.03 \\
            & Uba    \cite{kn:uba1}  &&                 &        &        1.60           & 0.30          &       \\
\cline{2-8}
            & Expt. Cable \cite{kn:kootte}     &        &          &    & 1.70          & 0.25          & 0.98 \\
            & Expt. van Laar\cite{kn:kootte}&           &               &    & 1.60          & 0.30          & 0.95 \\
\mr
0.75(L1$_2$)&vBH(this work)& 1.71          & 0.25         & 0.62  & 1.74          & 0.26         & 0.64 \\
            &VWN(this work)& 1.75          & 0.26         & 0.63  & 1.82          & 0.27         & 0.65 \\
            &LMH(this work)& 1.83          & 0.28         & 0.67  & 1.87          & 0.28         & 0.68 \\
            &PW(this work) & 1.95          & 0.36         & 0.76  & 1.82          & 0.31         & 0.69 \\
            & Auluck \etal\cite{kn:kashyap}&&               &        & 1.85           & 0.25          & 0.65 \\
            & Kootte \etal\cite{kn:kootte} &&               &        & 1.69           & 0.27          & 0.63 \\
            & Tohyama \etal\cite{kn:toh}   &&               &        & 2.88           & 0.38          &       \\
            & Lange \etal \cite{kn:lange}  & 1.72           & 0.25      & 0.62        &               &&       \\
            & Uba \etal   \cite{kn:uba1}  &&                 &        & 1.74           & 0.24          &       \\ 
\cline{2-8}
            & Expt. Menginger \cite{kn:kootte}       &       &               &        & 1.64                   & 0.26          & 0.61   \\
            & Expt. Lange \etal \cite{kn:lange}&&                          &        &                &               & 0.70  \\                                                      
\br
\end{tabular}
\end{center}
\label{tab4}
\end{table}

\subsection{Magnetism in Co-Pt Alloys}
\subsubsection{Ordered Alloys :}

Table 4 shows the calculated and experimental magnetic moments for ordered Co-Pt alloys.
No experimental result is  available for $25\%$ of concentration of Pt in ordered case. 
The local as well as average magnetic moments obtained by Auluck \etal \cite{kn:kashyap} using vBH
exchange correlation potential with experimental lattice parameter are lower  
(by 0.30 $\mu_B$ for Co site, 0.01 $\mu_B$ for Pt site and 0.23 $\mu_B$ for average) than 
our corresponding values which could be due to differences in computational 
details as mentioned in the case of Fe-Pt. 
The local as well as average magnetic moments obtained by 
Kootte \etal \cite{kn:kootte} using localized spherical wave method using vBH 
exchange correlation and experimental lattice parameters are in agreement with our corresponding 
values. 

For 50$\%$ concentration of Pt, our results agree well with the previous theoretical results 
\cite{kn:kashyap,kn:kootte,kn:uba1} within the errorbars of different calculational 
schemes and are in reasonable agreement with the observed magnetic moments \cite{kn:kootte} 
as summarized in Table 4.

For 75$\%$ concentration of Pt, the calculated local magnetic moments on Co site and that of 
average magnetic moments using possible exchange correlations with both theoretically 
estimated as well as experimental lattice parameters are on higher side as compared to the 
experimental estimates \cite{kn:kootte}. The calculated local moment of Pt using vBH 
exchange correlation and theoretically estimated lattice parameter is close to the experimental value \cite{kn:kootte}.   
The theoretical estimates for local as well as average magnetic moments by Auluck \etal \cite{kn:kashyap} 
and Kootte \etal \cite{kn:kootte} as in 50$\%$ concentration of Pt are in agreement with 
our corresponding estimates as can be seen from Table 4. The slight differences seen are 
again due to the differences in computational details. 
The local magnetic moments calculated by Tohyama \etal \cite{kn:toh} using tight binding method 
are significantly higher than ours as well as experimental estimates which can be seen from 
Table 4. The recent work by 
Lange \etal \cite{kn:lange} using fully relativistic TB-LMTO with vBH exchange 
correlation and theoretically estimated lattice parameter report the local as well as average 
magnetic moment close to our corresponding values. Their experimental value for average 
magnetic moment matches with our corresponding calculated value using  
non local exchange correlation potentials and experimentally estimated lattice parameter. The 
supercell calculation of Uba \etal \cite{kn:uba1} with LMTO using vBH exchange correlation potential 
and experimental lattice parameter matches well with our corresponding value.    

\subsubsection{Disordered Alloys :}

\n In Figure 3, we have shown the comparison of local magnetic moments of Co and Pt as well as average
magnetic moment of disordered Co-Pt system. Calculations have been done both within ASR and CPA schemes 
using vBH exchange correlations. 
The comparison with experimental results for average
magnetic moment taken from Landoldt series \cite{kn:expt} matches well with our calculations. The calculated
magnetic moments with augmented space recursion method are in better agreement with experimental
results than that of coherent potential approximation method. From this Figure we can see that the
local moment of Co  obtained by ASR calculation is almost constant with the increase of concentration of Pt 
which is the signature of weak local environmental effect on Co site. This finding 
is in agreement with that of Sanchez \etal \cite{kn:sanchez} who also 
pointed out almost constant magnetic moment at Co site as a function of Pt concentration. 
The average magnetic moments obtained by Koepernik \etal \cite{kn:kvhe} using linear combination of 
atomic orbitals combined with coherent potential approximation (LCAO-CPA)
method taking into account both diagonal and off diagonal disorder effects show close 
agreement with our results (except 20$\%$ concentration of Pt where the value obtained 
by Koepernik \etal \cite{kn:kvhe} is in higher side than ours) using augmented space 
recursion (ASR). 

\begin{figure}
\centering
\epsfxsize=4.0in\epsfysize=3.5in
\rotatebox{270}{\epsfbox{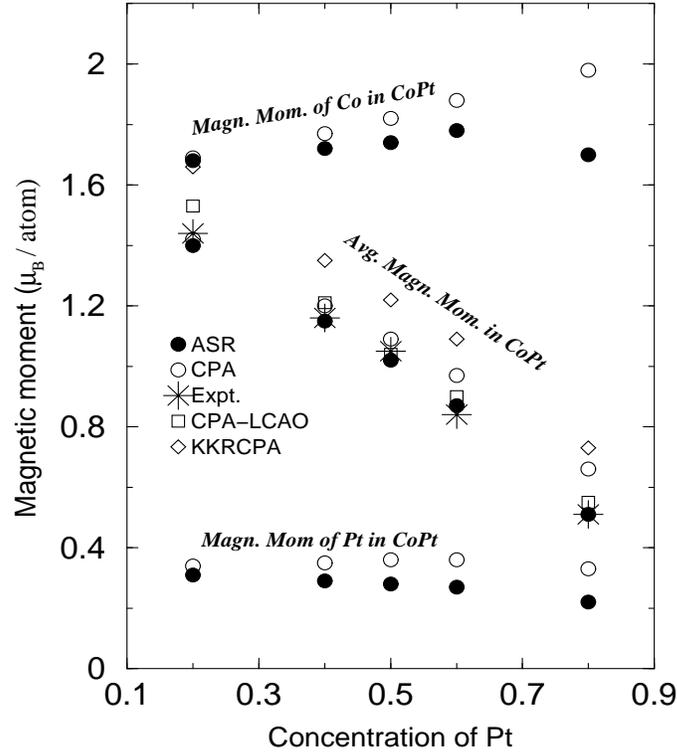}}
\caption{Magnetic moments in disordered Co-Pt alloy systems using two different linearized muffin-tin orbital (LMTO) 
based configuration averaging methods namely augmented space
recursion (ASR) and coherent potential approximation (CPA) as compared to available experimental values given in
Landoldt series \cite{kn:expt}. CPA-LCAO and KKR-CPA denote coherent potential approximation based linear combination of atomic 
orbitals method of Koepernik \etal \cite{kn:kvhe} and Korringa Kohn Rostoker coherent potential approximation 
method of Ebert \etal \cite{kn:ebert1} respectively.}
\label{fig3}
\end{figure}

The results obtained by Ebert \etal \cite{kn:ebert1} using Korringa-Kohn-Rostoker coherent 
potential approximation (KKR-CPA) are higher than ours as well as experimental values. The 
calculations by Ebert \etal \cite{kn:ebert1} using KKR-CPA with single site 
approximation were though fully relativistic did not take into account lattice relaxation 
and off diagonal disorder effects. Therefore it is not surprising that our calculations 
show better agreement with experiments. According to the calculation of Shick \etal \cite{kn:shick} 
using fully relativistic linearized muffin-tin orbital based coherent potential approximation 
(LMTO-CPA) method the average and partial magnetic moments of Co and Pt in Co$_{50}$Pt$_{50}$ are 
1.07, 1.79 and 0.35 $\mu_{B}$ respectively while the value of Ghosh \etal \cite{kn:gcsm} 
using ASR are 1.05, 1.85 and 0.24 $\mu_{B}$ for the same. Our values in this case are
1.05, 1.80 and 0.29. The reason behind the differences seen in between LMTO-CPA of 
Shick \etal \cite{kn:shick} and ASR is again same as explained above in the connection 
with KKR-CPA and ASR. Though the calculations by Ghosh \etal \cite{kn:gcsm} (using theoretically 
estimated lattice parameter) and ours (using experimental lattice parameter) used same ASR method, 
ours being charge neutral and self consistent show better agreement for local magnetic moments 
with corresponding charge neutral and self consistent calculations. 

In order to investigate the possible influence of short range order on the disordered magnetic moments, 
we have performed a complete investigation in terms of the total energy calculations as a
function of short range order parameter. Like Fe-Pt, Co-Pt also shows a tendency to order. 
We have also checked the variation of magnetic moments as a function of SRO parameter and find 
almost constant local as well as average magnetic moments as SRO parameter goes from segregating 
side to ordering side confirming the very little effect of short range order on magnetism of Co-Pt 
alloy system. 

\begin{table}
\caption{The local and average magnetic moments of Ni-Pt system in ordered structures with various 
choices of exchange correlation functionals.}
\begin{center}
\begin{tabular}{|c|l|c|c|c|c|c|c|}
\br
Concentration& XC used/Expt/ & \multicolumn{6}{|c|}{magnetic moment ($\mu_B$/atom) of}\\ \cline{3-8}
of Pt        & Ref.      &\multicolumn{3}{c|}{with eq. lat. par.} & \multicolumn{3}{c|}{with expt. lat. par.}      \\ \cline{3-8}
             &               & Ni     & Pt   &average & Ni     & Pt   &average\\ \cline{1-8}
\hline
0.00(FCC)&vBH(this work)& 0.61         &              &            & 0.62    &              &  \\
         &VWN(this work)& 0.62         &              &            & 0.64    &              &  \\
         &LMH(this work)& 0.64         &              &            & 0.65    &              &  \\
         &PW(this work) & 0.66         &              &            & 0.64    &              &  \\
\cline{2-8}
         & Expt. \cite{kn:crc}   & 0.62         &              &            &          &              &  \\
\hline
0.25(L1$_2$)&vBH(this work)& 0.50         & 0.24        & 0.43      & 0.57   & 0.27        & 0.49 \\
            &VWN(this work)& 0.54         & 0.26        & 0.47      & 0.60   & 0.29        & 0.52 \\
            &LMH(this work)& 0.62         & 0.29        & 0.53      & 0.65   & 0.30        & 0.56 \\
            &PW(this work) & 0.71         & 0.36        & 0.63      & 0.63   & 0.32        & 0.56 \\
            & Singh \cite{kn:singh}      & 0.58      & 0.27         & 0.50   &             & &       \\
\cline{2-8}
            & [Expt.     &           &              &                        & 0.49        & 0.25   & 0.43     \\
            & Parra \etal]\cite{kn:parra}&           &              &            &          &              &       \\
\hline
0.50(L1$_0$)&vBH(this work)    & 0.00        & 0.00        & 0.00       & 0.33   & 0.17        & 0.25 \\
            &VWN(this work)    & 0.06        & 0.03        & 0.05       & 0.46   & 0.23        & 0.34 \\
            &LMH(this work)    & 0.55        & 0.27        & 0.41       & 0.65   & 0.31        & 0.48 \\
            &PW(this work)     & 0.72        & 0.34        & 0.53       & 0.63   & 0.32        & 0.48 \\
            & Singh \cite{kn:singh}& 0.60    & 0.27        & 0.44       &        &             &       \\
\cline{2-8}
            & [Expt            &              &              &          & 0.28   & 0.17        & 0.22 \\
            & (Parra \etal)] \cite{kn:parra}   &              &              &             &         &              &       \\
\hline
0.75(L1$_2$)&vBH(this work)    & 0.47        & 0.09        & 0.18       & 0.55   & 0.10        & 0.21 \\
            &VWN(this work)    & 0.50        & 0.09        & 0.20       & 0.57   & 0.11        & 0.22 \\
            &LMH(this work)    & 0.55        & 0.11        & 0.22       & 0.61   & 0.12        & 0.24 \\
            & PW(this work)    & 0.65        & 0.16        & 0.28       & 0.58   & 0.12        & 0.24 \\
            & Singh \cite{kn:singh}& 0.58     & 0.10       & 0.22       &         &            &       \\                
\hline
\end{tabular}
\end{center}
\label{tab6}
\end{table}

\subsection{Magnetism in Ni-Pt Alloys}
\subsubsection{Ordered Alloys :}

In Table 5, we show two sets of ordered calculations in Ni-Pt alloys using possible 
local as well as non local exchange correlation potentials one with theoretically calculated lattice constants 
via energy minimization procedure and other using experimental lattice parameters. 

For 25$\%$ concentration of Pt, the calculated local as well as average magnetic moments in ordered Ni-Pt 
alloys obtained using vBH local exchange correlation potential in theoretically 
calculated lattice parameter show very good agreement with experimental values \cite{kn:parra}.  
The values obtained by Singh \cite{kn:singh} in the same case are higher in comparison to ours 
and experimental estimate \cite{kn:parra}.
Singh's \cite{kn:singh} calculations seemingly did not include the $f$ states in Pt in the TB-LMTO basis. Our test 
calculations without including $f$ states of Pt also show higher values of magnetic moments 
for this concentration of Ni-Pt alloy. 

For 50$\%$ concentration of Pt in L1$_0$ structure
calculated local as well as average magnetic moments using vBH exchange correlation 
potential with the use of experimental lattice parameter is closest to the experimental estimate \cite{kn:parra}. Our 
calculations with the use of local exchange correlations and theoretically estimated lattice 
parameters lead to non magnetic ground state which is in agreement with that found in previous study by 
Dahmani \etal \cite{kn:dahmani}. In our calculations we have taken into account the tetragonal distortion as in the case of 
Fe-Pt and Co-Pt alloys in L1$_0$ structure.

For 75$\%$ concentration of Pt, for NiPt$_3$ alloy in L1$_2$ structure there is no experimental result available. 
For this concentration we
have got higher local magnetic moment of Ni than at the 50$\%$ concentration of Pt. This was obtained while
using local exchange correlations. In this case if we use non local exchange correlations then we get the decrease of local
magnetic moment of Ni on going from 50$\%$ to 75$\%$ concentration of Pt. The average as well 
as local magnetic moments on Pt sites show the decreasing tendency using both local as well as 
non local exchange correlations with theoretically as well as experimentally estimated lattice constants. 

The calculations by Singh \cite{kn:singh} using vBH exchange correlations and theoretically estimated 
lattice parameters show that the local magnetic moment of Ni increases while going from 25$\%$ to 50$\%$ and 
decreases while going from 50$\%$ to 75$\%$ concentration of Pt. The calculations by Singh \cite{kn:singh} did not take
into account the tetragonal distortion for 50$\%$ concentration of Pt which means putting lattice parameters 
$a = c$ which is not the right ground state structure. For a test we also repeated our calculation without taking into account
the tetragonal distortion for 50$\%$ concentration of Pt using vBH local exchange correlation potentials 
and theoretically estimated lattice constants and we also observed same trend as Singh obtained. 
However, for the calculation taking into account the degrees of freedom for tetragonal distortion we found
that the magnetic moments vanish with the use of local exchange correlation potentials in theoretically estimated 
lattice parameters.

\begin{figure}
\centering
\epsfxsize=6.5in\epsfysize=2.5in
\rotatebox{0}{\epsfbox{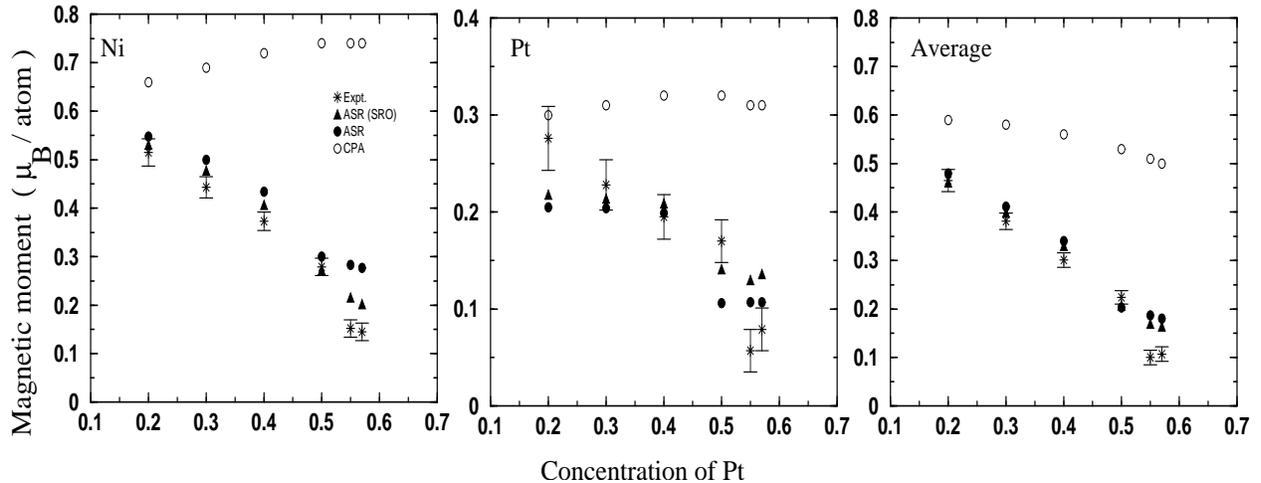}}
\caption{Magnetic moments in disordered Ni-Pt alloy systems using two different configuration averaging 
methods namely augmented space recursion (ASR) and coherent potential approximation (CPA) as compared to 
experimental values given by Parra \etal \cite{kn:parra}. ASR (SRO) denotes the results with short range ordering 
effect using ASR.}
\end{figure}

\subsubsection{Disordered Alloys :}
We have plotted the local and average magnetic moments of disordered Ni-Pt system in Figure 4. The comparison of calculated 
disordered magnetic moments using augmented space recursion (ASR) method with vBH exchange correlation potentials 
and experimental lattice parameter matches 
well with experimental values \cite{kn:parra} in all concentrations except $55\%$ and 
$57\%$ of Pt. Our calculations of magnetic moments using coherent potential approximation (CPA) 
method using vBH exchange correlation potential 
and experimental lattice parameters are very different than the calculations using ASR method and 
experimental estimates \cite{kn:parra}.
Using CPA the local magnetic moments of Ni donot even follow the trend of 
corresponding experimental estimates. ASR being capable to go beyond single site approximation 
taking into account lattice relaxation and off diagonal disorder effect which is very important 
in NiPt alloys as was shown in our previous paper \cite{kn:dta} provides better agreement with 
experiment than CPA. Our calculated values for $55\%$ and $57\%$ concentration of Pt using ASR 
method are in higher side in comparison to the experimental estimates which leads us to suspect 
the presence of short range ordering effect. We performed calculations incorporating short range 
order for all concentration of Pt in this system and found that the magnetic moments of Ni 
decreases by appreciable fraction for $55\%$ and $57\%$ concentration of Pt. The moment of Pt increases 
slightly. These give rise to the decrease of average magnetic moment in these concentrations. 
Calculations incorporating the effect of short range order agrees well with experimental estimate 
of parra \etal \cite{kn:parra}.

\section{Summary and conclusions}

To summarize, our study for ordered alloys to investigate the role played by 
different possible exchange correlation functionals shows that choice of the 
exchange-correlation potential has considerable effect on the values of the 
equilibrium lattice constants as well as magnetic moments. 

The present study on disordered alloys shows that the single site approximation
based methods work reasonably well for  Fe-Pt
systems and is in close agreement with our ASR predictions. 
For the Co-Pt system, the CPA begins to deviate from the ASR. 
CPA based calculations  show slight increase in 
the local magnetic moment of Co with increasing Pt concentration, while the ASR  
shows almost constant behaviour. This indicates  the signature of weak local environmental
effect on Co sites. 

It is in the Ni-Pt alloy that CPA shows the largest deviation from the ASR. 
The CPA estimates of  the magnetic moments are quite different from the 
experimental values. It predicts increase of the local magnetic moment on Ni
with increasing Pt concentration, whereas experimentally the reverse behaviour
is observed. In the absence of local environment effects, increase of Pt
concentration in Ni-Pt should lead to increase in the local Ni moment, since
isolated clusters of Ni in Pt become more probable. This leads to narrowing of the
local density of states on Ni and consequently according to the Stoner picture,
an increase in the local Ni moment. Finally in the dilute limit, this local moment should
approach the moment of an isolated  Ni atom. This behaviour is certainly seen in Fe-Pt alloys. 
However, the fragile moment on Ni seems to need at least 50$\%$ Ni atoms in its nearest neighbour 
environment, otherwise it loses its local moment. This is indeed what one sees in experiment and 
is a strong indicator of large local environmental effect in Ni-Pt. 
The CPA predicts increase of the local magnetic moment on Ni
with increasing Pt concentration. This is expected, since the CPA does not take into
account the effect of local environment. The ASR, however, predicts the correct trend with
increasing Pt concentration. The estimates of the actual value of the local magnetic moments
are also much better.

Our total energy calculation as a function of short range 
order confirms the ordering tendency in these systems. The calculation of magnetic 
moments as a function of short range order shows that its effect is small on 
the magnetism in  Fe-Pt and Co-Pt disordered alloys but significant 
on the magnetism of  disordered Ni-Pt.  

Finally, the numerical details of calculations, convergence with the number of k-points in the
Brillouin zone integrations, choice of atomic sphere radii, proper convergences in the
CPA and the ASR, the proper choice of the minimal basis set in the TB-LMTO, all of these affect 
the actual values of the estimated magnetic moments. 

\appendix
\section{}
Details of  the methodology of augmented space recursion has been presented in an earlier papers referred in the text. 
Here we shall quote the key results of TBLMTO-ASR generalized to take in account the short-range ordering
effect. The augmented space Hamiltonian including short range order can be written as

\begin{eqnarray}
\^H = H_{1} + H_{2}\sum_{R} P_{R} \otimes P_{\downarrow}^{R} +
H_{3}\sum_{R} P_{R} \otimes (T_{\downarrow\uparrow}^{R}+T_{\uparrow\downarrow}^{R}) \nonumber\\
 +H_{4}\sum_{R}\sum_{R'}T_{RR'}\otimes\unit+ \alpha H_{2}\sum_{R''} P_{R''}\otimes P_{\downarrow}^{1}
\otimes (P_{\uparrow}^{R''}-P_{\downarrow}^{R''}) \nonumber\\
 +H_{5}\sum_{R''} P_{R''}\otimes P_{\downarrow}^{1}\otimes (T_{\uparrow\downarrow}^{R''}+T_{\downarrow\uparrow}^{R''}) \nonumber\\
 +H_{6}\sum_{R''} P_{R''}\otimes P_{\downarrow}^{1}\otimes (T_{\uparrow\downarrow}^{R''}+T_{\downarrow\uparrow}^{R''}) \nonumber\\
 +\alpha H_{2} \sum_{R''} P_{R''}\otimes(T_{\uparrow\downarrow}^{1}+T_{\downarrow\uparrow}^{1})\otimes
(P_{\uparrow}^{R''}-P_{\downarrow}^{R''}) \nonumber\\
 + H_{7}\sum_{R''} P_{R''} \otimes (T_{\uparrow\downarrow}^{1}+T_{\downarrow\uparrow}^{1})\otimes
(T_{\uparrow\downarrow}^{R''}+T_{\downarrow\uparrow}^{R''})
\end{eqnarray}

\n where $R''$ belong to the set of nearest neighbours of the site labeled $R$, at which the local density
of states will be calculated. $P$'s and $T$'s are the projection and transfer operators either in the
space  spanned by the tight-binding basis $\{\vert R\rangle\}$ or the configuration space associated with the sites
$R$ spanned by  $\{\vert \uparrow_R\rangle, \vert\downarrow_R\rangle \}$ as described in
\cite{kn:ppb}. 

\[ \vert \uparrow_R\rangle = \sqrt{x}\vert A_R\rangle+\sqrt{y}\vert B_R\rangle\quad\quad
 \vert \downarrow_R\rangle = \sqrt{y}\vert A_R\rangle-\sqrt{x}\vert B_R\rangle\]

\n The different terms of the Hamiltonian are given below.

\begin{eqnarray}
H_{1} = A(C/\Delta)\Delta_{\lambda} -(EA(1/\Delta)\Delta_{\lambda}-1) \nonumber\\
H_{2} = B(C/\Delta)\Delta_{\lambda} -EB(1/\Delta)\Delta_{\lambda} \nonumber\\
H_{3} = F(C/\Delta)\Delta_{\lambda} -EF(1/\Delta)\Delta_{\lambda} \nonumber\\
H_{4} = (\Delta_{\lambda})^{-1/2} S_{RR'}(\Delta_{\lambda})^{-1/2} \nonumber\\
H_{5} = F(C/\Delta)\Delta_{\lambda} [\sqrt{(1-\alpha)x(x+\alpha y)} + \sqrt{(1-\alpha)y(y+\alpha x)} -1] \nonumber\\
H_{6} = F(C/\Delta)\Delta_{\lambda} [y\sqrt{(1-\alpha)(x+\alpha y)/x} + x\sqrt{(1-\alpha)(y+\alpha x)/y} -1] \nonumber\\
H_{7} = F(C\Delta)\Delta_{\lambda} [\sqrt{(1-\alpha)y(x+\alpha y)} - \sqrt{(1-\alpha)x(y+\alpha x)} \\
\mathrm{where}\nonumber\\
A(Z)=xZ_{A}+yZ_{B} \nonumber\\
B(Z)=(y-x)(Z_{A}-Z_{B}) \nonumber\\
F(Z)=\sqrt{xy}(Z_{A}-Z_{B})\nonumber
\end{eqnarray}

\n $\alpha$ is the nearest neighbour Warren-Cowley parameter described earlier. $\lambda$ labels  the constituents
A or B in case of binary AB alloy.
$C$'s and $\Delta$'s are the potential parameters describing the atomic scattering properties of the constituents
 and $S$ is the screened structure constant describing the
underlying lattice which is face-centered cubic (FCC) in the present case. For convenience, all the angular
momentum labels have been suppressed, with
the understanding that all potential parameters are 9$\times$9 matrices (for an $spd$ minimal  basis set).
We note that in absence of short-ranged order ($\alpha$ = 0), the terms
H$_{5}$ to H$_{7}$ disappear and the Hamiltonian reduces to the standard one described earlier \cite{kn:ppb}.


\section*{References}
\begin{thebibliography}{99}
\bibitem{kn:cad} Cadeville M C and Mor\'an-L\'opez J L 1987 {\it Physics Reports} {\bf 153} 331
\bibitem{kn:mir} Mirebeau L., Cadeville M.C., Parette G. and Campbell I.A., 1982 \JPF {\bf 12} 25
\bibitem{kn:cad2} Pierron-Bohnes V.,Cadeville M.C. and Parette G., 1985 \JPF {\bf 15} 1441
\bibitem{kn:mir1} Mirebeau I.,Hennion M. and Parette G., 1985 \PRL {\bf 53} 687
\bibitem{kn:cad3} Pierron-Bohnes V.,Cadeville M.C.  and Gautier F., 1983 \JPF {\bf 13} 1689
\bibitem{kn:sato} Sato H., Arrott A. and Kikuchi R., 1959 {\it Journal of Physics and Chemistry of solids}
{\bf 10} 19
\bibitem{swa} Swalin R.A., {\it Thermodynamics of solids},Wiley,New York, 1962
\bibitem{von} Vonsovskii S.V., {\it Magnetism } (Wiley,New York,1974)
\bibitem{kn:bieber} Bieber A., Gautier F.,Treglia G. and Ducastelle F., 1981 {\it Solid State Commun} {\bf 39}
149
\bibitem{bieber1} Bieber A. and Gautier F., 1981 \SCC {\bf 38} 1219
\bibitem{kn:bieber2} Bieber A. and Gautier F., 1986 {\JMMM} {\bf 54 -57} 967
\bibitem{kn:hen} Hennion M., 1983 \JPF {\bf 13} 2351
\bibitem{kn:jac} Jaccarino V. and Walker J.L., 1965 \PR {\bf 15} 258
\bibitem{kn:mar} Marshall W., 1968 \JPC {\bf 1} 88
\bibitem{kn:hicks}Hicks T.J., 1970 {\it Physics Letters} {\bf A32} 410
\bibitem{kn:has} Hasegawa H. and Kanamori J., 1971 {\it J. Phys. Soc. Japan} {\bf 31} 382
\bibitem{kn:but} Buttler W.H., 1973 \PR {\bf B8} 4499
\bibitem{kn:jo} Jo T. amd Miwa H., 1976 {\it J. Phys. Soc. Japan} {\bf 40} 706,
\bibitem{kn:jo2} Jo T., 1976 {\it J. Phys. Soc. Japan} {\bf 40} 715
\bibitem{kn:has2} Hasegawa H., 1979 {\it J. Phys. Soc. Japan} {\bf 46} 1504
\bibitem{kn:ham} Hamada N.,  1979 {\it J. Phys. Soc. Japan} {\bf 46} 1759
\bibitem{kn:kak} Kakehashi Y. 1982 {\it J. Phys. Soc. Japan} {\bf 51} 94
\bibitem{kn:uba} Uba S. {\it et al} 1998 \PR {\bf B57} 1534; Geerts W. 1994  {\it et al} \PR {\bf B50} 12581; Weller D.,Harp G.R.,
Farrow R.F.C.,Cebollada A. and Sticht J. 1994 \PRL {\bf 72} 2097
\bibitem{kn:asr} Saha T., Dasgupta I. and Mookerjee A., 1996 \JPCM {\bf 8} 1979
\bibitem{kn:mook} Mookerjee A., 1973 \JPC {\bf 6} L205 
\bibitem{kn:hhk} Haydock R.,Heine V. and Kelly M.J., 1972 \JPC {\bf 5} 2845 
\bibitem{kn:vol35} Haydock R., {\it Solid State Physics} {\bf 35}
(Academic Press, N. Y. ) (1988)
\bibitem{kn:sm} Chakrabarti A and Mookerjee A., \JPCM {\bf 13} 10149(2001)
\bibitem{kn:latdis} Saha T. and Mookerjee A., \JPCM {\bf 8} 2915 (1996b)
\bibitem{kn:zun} Lu Z.W.,Laks D.B.,Wei S.H. and Zunger A., 1994 \PR {\bf B50} 6642
\bibitem{kn:staun} Staunton J.B.,Johnson D.D. and Pinski F.J., 1994 \PR {\bf B50} 1450
\bibitem{kn:john} Johnson D.D.,Staunton J.B. and Pinski F.J., 1994 \PR {\bf B50} 1473
\bibitem{kn:borici} Borici-Kuqo M., Monnier R. and Drchal V., 1998 \PR {\bf B58} 8355
\bibitem{kn:zun2} Wolverton C., Ozolins V.  and Zunger A. 1998 \PR {\bf B57} 4332
\bibitem{kn:abri} Abrikosov I. A., Niklasson A. M. N.,  Simak S. I., and Johansson B. 1996 \PRL {\bf 76} 4203
\bibitem{kn:cow} Cowley J.M., 1950 {\it J. Appl. Phys.} {\bf 21} 24
\bibitem{kn:as} Mookerjee A. and Prasad R., 1993 \PR {\bf B48} 17724
\bibitem{kn:saha} Saha T.,Dasgupta I. and Mookerjee A. 1994 \PR {\bf B50} 13267
\bibitem{kn:gcsm} Ghosh S., Chaudhuri C. B., Sanyal B. and Mookerjee A., 2001 \JMMM {\bf 234} 100
\bibitem{kn:vbh} von Barth U. and Hedin L., 1972 \JPC {\bf 5} 1629
\bibitem{kn:vwn} Vosko S. H., Wilk L. and Nusair M., 1980 {\it Can. J. Phys.} {\bf 58} 1200
\bibitem{kn:lmh} Langreth D. C. and Mehl M. J., 1981 \PRL {\bf 47} 446
\bibitem{kn:pw} Perdew J. P. and Wang Y., 1986 \PR {\bf B33} 8800
\bibitem{kn:ajs} Andersen O.K. and Jepsen O. 1984 \PRL {\bf 53} 2581
\bibitem{kn:oka} Andersen O.K., Jepsen O. and \v{S}ob, {\sl Electronic Band Structure and Its Applications}
  ed. M. Yussouff. Lecture Notes in Physics 283, Springer 1987 , 1992.
\bibitem{kn:oja} Andersen O.K., Jespsen O. and Krier G., {\sl Lectures on Methods of
Electronic Structure Calculations} eds.: V. Kumar, O. K. Andersen, A. Mookerjee. Singapore, World
  Scientific , 1994.
\bibitem{kn:ddsam} Das G.P., {\sl Electronic Structure of Alloys, Surfaces and Clusters}, Advances in
Condensed Matter Science, Vol. {\bf 4,} eds.: A. Mookerjee and D. D. Sharma, Taylor-Francis, 2003
\bibitem{kn:bs} Sanyal B., Biswas P.P., Mookerjee A., Das G.P., Salunke H. and Bhattacharya A.K.,
1998 \JPCM {\bf 10} 5767
\bibitem{kn:ppb} Biswas P.P., Sanyal B., Fakhruddin M., Halder A., Mookerjee A. and Ahmed M.,
1995 \JPCM {\bf 7} 8569
\bibitem{kn:kd} Kudrnovsk\'y J. and Drchal V., 1990 \PR {\bf B41} 7515
\bibitem{kn:gdm} Ghosh S., Das N. and Mookerjee A., 1999  \JPCM {\bf 9} 10701
\bibitem{kn:sdgthesis} Ghosh S.D., Ph.D. Thesis, Jadavpur University, (2000)
\bibitem{kn:pearson} Pearson W.B., {\sl A handbook of lattice spacings and structures of metals and
  alloys}  Oxford, Pergamon Press 1958-1967
\bibitem{kn:kashyap} Kashyap A., Garg K. B., Solanki A. K., Nautiyal T. and Auluck S., 1999 \PR {\bf B60} 2262
\bibitem{kn:pod} Podg\'orny M., 1991 \PR {\bf B43} 11300 ; 1992 \PR {\bf B46} 6293
\bibitem{kn:pisanty} Pisanty A., Amador C., Ruiz Y. and de la Vega M., 1990 {\it Z. Phys. B-Condensed Matter} {\bf 80} 237
\bibitem{kn:crc} Lide D.R.,  {\it Hand Book of Chemistry and Physics} 81st Ed. p12-119 (2000)
\bibitem{kn:ost} Osterloh I., Oppeneer P. N., Sticht J. and Kubler J., 1994 \JPC {\bf 6} 285
\bibitem{kn:toh}  Tohyama T.,  Ohta Y.  and  Shimizu M., 1989 \JPC {\bf 1} 1789 
\bibitem{kn:kky}  Kulikov N. I., Kulatov E.T. and Yakhimovich S.I., 1985 \JPF {\bf 15} 1127; 
\bibitem{kn:kub} Uhl M., Sandratskii L.M., and Kubler J., 1994 \PR {\bf B50} 291
\bibitem{kn:expt} Wijn H. P. J., Landolt-Bornstein (Eds.), {\it Magnetic Properties of Metals, 3d, 4d and 5d Elements, 
Alloys and Compounds}, New Series, Group III, Vol. 19 Part a, Springer-Verlag, Berlin, 1986
\bibitem{kn:hd} Hayn R. and Drchal V. (1998) \PR {\bf B58} 4341
\bibitem{kn:ito} Ito Y., Sasaki T. and Mizoguchi T., 1974 {\it Solid State Commun.} {\it 15} 807
\bibitem{kn:capo} Caporaletti O. and Graham G. M., 1980 \JMMM {\bf 22} 25
\bibitem{kn:stau} Major Z.S., Dugdale S., Jarlborg B., Bruno E., Ginatempo B., Staunton J.B. and Poulter J., 2003
\JPC {\bf 15} 3619
\bibitem{kn:kvhe} Koepernik K., Velick\'y B., Hayn R., and Eschrig H., 1997 \PR {\bf B55} 5717
\bibitem{kn:kootte}  Kootte A.,  Haas C. and  de Groot R.A.,  1991  \JPC {\bf 3} 1133
\bibitem{kn:uba1} Uba L., Uba S., Antonov V. N., Yaresko A. N., and Gontarz R,  2001  \PR {\bf  B64} 125105 
\bibitem{kn:lange} Lange R. J., Lee S. J., Lynch D. W., Canfield P. C., Harmon B. N. and  Zollner  S.,
1998 \PR {\bf  B58} 351       
\bibitem{kn:sanchez} Sanchez J.M. and Mor\'an-L\'opez, 1989 \JPCM {\bf 1} 491
\bibitem{kn:ebert1} Ebert H., Drittler B. and Akai H., 1992 {\it Journal of Magnetism and Magnetic Materials} {\bf 104-107} 733
\bibitem{kn:shick} Shick A. B., Drchal V., Kudrnovs\'ky J. and Weinberger P., 1996 \PR {\bf B54} 1610
\bibitem{kn:singh} Singh P. P., 2003 \JMMM {\bf 261} 347
\bibitem{kn:parra} Parra R. E. and Cable J. W., 1980 \PR {\bf B21} 5494
\bibitem{kn:dahmani} Dahmani C.E., Cadeville, M.C., Sanchez J.M. and Mor\'an-L\'opez J. L., 1985 \PRL {\bf 55} 1028
\bibitem{kn:dta} Paudyal D., Saha-Dasgupta T. and Mookerjee  A., 2003 \JPCM {\bf 15} 1029 
\end{thebibliography}
\end{document}